\input epsf
 
\magnification= \magstep1  
\tolerance=1600 
\parskip=6pt 
\baselineskip= 6 true mm

\font\medrm=cmr9  		

\def\a{\alpha}		\def\b{\beta}	
\def\g{\gamma} 		
\def\d{\delta} 		\def\D{\Delta}	\def\e{\varepsilon}
\def\h{\eta}			
 	 
\def\m{\mu}		\def\f{\phi} 	
		\def\vv{\varphi}
\def\n{\nu}		\def\j{\psi} 	
		\def\s{\sigma} 	
		\def\th{\theta}	
 	
\def\x{\xi}

 \def\LL{{\cal L}} 

\def\cl{\centerline} 	\def\ni{\noindent}
\def\pa{\partial}	\def\dd{{\rm d}}	 
\def\ra{\rightarrow}

\def\fnd#1{\footnote{$^\dagger$} {\scrunch #1 \toe}}
\def\fndd#1{\footnote{$^\ddagger$} {\scrunch #1 \toe}}
	\def\scrunch{\baselineskip=10 pt \medrm}
	\def\toe{\hfil\break\vskip-18pt}
	\newcount\noteno
\def\numfn#1{\global\advance\noteno by 1 
	\footnote{$^{\the\noteno}$} {\scrunch #1 \toe}}
 	
\def\ddt{{{\rm d}\over {\rm d}t}}
\def\fract#1#2{{\textstyle{#1\over#2}}}
\def\half{\fract12} \def\quart{\fract14}

\def\part#1#2{{\partial#1\over\partial#2}} 
 \def\ref#1{${\vphantom{)}}^{#1}$}

\def\qu{\ {\buildrel {\displaystyle ?} \over =}\ }
 \def\ket{\rangle}

{\nopagenumbers

\vglue 1truecm
\rightline{THU-96/02}
\rightline{gr-qc/9601014}
\vfil
\cl{\bf QUANTIZATION OF POINT PARTICLES} 
\vskip .3 truecm
\cl{\bf IN 2+1 DIMENSIONAL GRAVITY}
\vskip .3 truecm
\cl{\bf AND SPACE-TIME DISCRETENESS} 
\vfil
\cl{G. 't Hooft }
\vskip 1 truecm
\cl{Institute for Theoretical Physics}
\cl{Utrecht University , P.O.Box 80 006}
\cl{3508 TA Utrecht, the Netherlands}
\vfil
\ni{\bf ABSTRACT}

By investigating the canonical commutation rules for gravitating
quantized particles in a 2+1 dimensional world it is found that these
particles live on a space-time lattice. The space-time lattice points
can be characterized by three integers.  Various representations are
possible, the details depending on the topology chosen for
energy-momentum space. We find that an $S_2\times S_1$ topology yields
a physically most interesting lattice within which first quantization
of Dirac particles is possible. An $S_3$ topology also gives a lattice,
but does not allow first quantized particles.

\vfil\eject}\pageno=2

\ni{\bf 1. INTROUCTION}\medskip

Reducing the number of space dimensions from three to two in quantum
gravity is a quite severe simplification that drastically changes the
physics one attempts to describe. Due to the absence of local dynamical
degrees of freedom Einstein's equations for gravity become almost
trivial: in the absence of matter space-time is featureless (flat, if
there is no cosmological constant). Hence pure gravity is a topological
theory in this case, and it has been investigated by a number of
authors\ref1.

Pure 2+1 dimensional gravity is of interest if one wishes to contemplate
cosmologies with highly non-trivial boundary conditions in the
space-time coordinates. However if one wishes to gain insights in local
aspects of gravity these boundary conditions are not so important. The
theory then becomes non-trivial only if one adds matter. One of the
simplest forms of matter a particle physicist can imagine is a single,
non-interacting scalar field with the following Lagrangian:

$$\LL_{\rm matter}\,=\,\sqrt{-g}(-\half
g^{\m\n}\pa_\m\f\pa_\n\f-\half m^2\f^2)\,.\eqno(1.1)$$ 

\ni In the absense of gravity the quantum version of this theory
generates a Hilbert space spanned by all $N$~particle states, where
each particle has a given momentum, and the particles do not interact.

What gravity does to {\sl classical\/} point particles in 2+1
dimensions is also clear\ref2: each particle cuts a cone in 2-space, as
seen from its own Lorentz rest frame. Since the surrounding space-time
is completely flat this system is still exactly solvable, although in
the $N$~particle case the book-keeping of the transitions after a long
stretch of time can become infinitely complicated and chaotic phenomena
can easily occur\ref3.

Does this suggest that the quantum mechanical version of the
gravitating point particle theory should also be soluble? Many
classically integrable systems are also quantum mechanically
integrable. But it appears that we can compare the classical theory
best with that of classical particles in a triangular box with
arbitrary angles, which is classically integrable only for finite time
intervals, but at large times chaotic, and quantum mechanically not
exactly solvable.

Actually, the situation for our toy model is much worse than for the
particle in a triangular box: we do not really know how to {\sl
define\/} the corresponding quantum model. Is it a functional
integral?  If we try to define the theory from that angle we must face
the fact that it is not renormalizable, so that one must seriously
doubt whether the functional integral can be at all meaningfully
defined. Should we build a Hilbert space from states with $N$ (moving)
punctures in space and time?\ref4  But that is a {\sl first quantized\/}
theory. If we want a theory that in the absense of matter would
approach a scalar field theory we want something that is {\sl second\/}
quantized. This would require the introduction of creation and
annihilation operators. Creating or annihilating punctures is tricky if
we want to use coordinates to localize these operators, since they
backfire on the coordinates themselves.

A promising avenue may be first to construct a Hilbert space containing
$N$ Dirac~like gravitating particles, obeying Fermi-Dirac statistics or
at least some sort of exclusion principle. This would be the first
quantized model. Then one might hope to be able to construct states
where the Dirac sea of negative-energy states is filled, and introduce
second quantization that way. This programme has not yet been carried
out completely. A problem is that the negative-energy particles in the
Dirac sea would generate negative curvature which should somehow have
to be compensated, perhaps by a cosmological constant.

In spite of all these obstacles it is of importance to try to find our
way in this jungle. If a meaningful model can be constructed it might
show us on the one hand how gravity takes care of its own
renormalization problems in the ultraviolet region, and on the other
hand, in the infrared region,  how to handle the system as a {\sl
quantum cosmology},  which requires a definition of cosmological time,
and a way to handle the measurement problem.

It has been noted before that the {\sl time\/} coordinate must probably
be chosen to be a discrete multiple of a Planckian time unit\ref5. This
important observation follows from the fact that the Hamiltonian is an
angular variable, well defined only {\sl modulo\/} $2\pi$ (in
conveniently chosen Planck units). Time is also directly linked to
angular momentum. But what about the spacelike coordinates?  This
author's first attempts to identify appropriate variables for the
{$N$-particle} Hilbert space did not lead to space quantization\ref5.
This was because we used the so-called polygon
representation\ref{3,5,6,7}.  Natural parameters to characterize an
{$N$-particle} state appeared to be the lengths $L_i$ of the edges of
polygons used to tesselate a two-dimensional Cauchy surface. The angles
of the polygons are fixed as soon as the Lorentz boost parameters are
given; that is, the Lorentz boosts $\eta_i$  relating the coordinate
frame of one polygon to that of a neighboring polygon to which it is
attached. At each $L_i$ we have one $\eta_i$, and it was discovered
that the $L_i$ and $\eta_i$ are conjugated variables:

$$\{2\eta_i,L_j\}\,=\,\d_{ij}\,.\eqno(1.2)$$

But the ``momenta'' $\eta_i$ are not angles, they are {\sl
hyperbolic\/} angles.  The equations of motion relate the trigonometric
functions of the Hamiltonian, $\sin H$ and $\cos H$, to the hyperbolic
sines and cosines of the $\eta$ parameters. If anything, this would
suggest that the imaginary parts, not the real parts, of the parameters
$L_i$ are to be chosen discrete.  If the $L_i$ were to represent the
space coordinates, they still appeared to live in a continuum.

But using the poygon representation to formulate Hilbert space leads to
other problems as well. We have very complicated boundary conditions in
the form of monodromies as soon as particles make a full swing around
each other.  Also there are constraints: each polygon must be a closed
polygon, which appears to correspond to the constraint that the state
must be invariant under Lorentz transformations of the coordinate
frames inside each polygon. The fact that the exterior angles of each
polygon must add up to $2\pi$ corresponds to the constraint of
invariance with respect to time boosts for the frame inside each
polygon. We found that these constraints are easier to deal with if
instead of the $L_i$ we use coordinates $\{x_i,y_i\}$ to localize each
corner of each polygon in its own frame. This required to deduce the
equations of motion and the Poisson brackets from scratch once again.

It is here that we hit upon a pleasant surprise: the momenta conjugated
to these $x_i$ and $y_i$ are true angles once again, as if space itself
were limited to discrete numbers as well. Is space as well as time a
lattice? If this is so it will be hard to recover rotational and
Lorentz invariance. Our first attempts were to replace the wave
equations by lattice equations which reduce Poincar\'e invariance to
one of its discrete subgroups. But this seriously jeopardized the
entire structure of the theory. If our monodromies were to be
restricted to lattice group elements it would become impossible to
consider continuum limits. This wasn't the theory we were looking for.

The present paper explains how to set up a more satisfactory scheme.
Complete invariance under the continuum Lorentz group needs not be 
sacrificed even if space and time are restricted to points on a lattice.
Everything is derived from the Poisson algebra, which leads one to 
conclude that the results are inevitable. It is somewhat puzzling 
to the present author why the lattice structure of space and time 
had escaped attention from other investigators up till now. 

There are different possible representations of the lattice. We first 
thought that the $S_3$ topology for energy-momentum space would be
the most natural choice. It will turn out however that in the lattice 
thus obtained no Lorentz-invariant quantum particles can exist unless
negative probabilities are accepted. Even a Dirac square root of 
the partial difference equations for the wave functions does not allow 
for a positive probability interpretation. We will demonstrate that 
this disease can be removed if we use an $S_2\times S_1$ topology 
instead. The lattice looks only slightly different, and the Dirac 
equation becomes a beautiful first order difference equation.
\bigbreak

\ni{\bf 2. THE EQUATIONS FOR ONE PARTICLE}\medskip

For a particle at rest, surrounding space is a cone with a deficit angle
$\b$ that can be identified with the mass $m$.  When the particle moves
it is convenient to consider the relation between {\sl half\/} the
deficit angles at motion and at rest. Therefore we will use units such
that the deficit angle, which is easily seen to be additively
conserved, is identified with twice the Hamiltonian $H$ of the system.
We will use the symbol $\m$ for half the deficit angle at rest, being a
parameter that corresponds to mass of a particle. For the time being we
will assume that $\m<\pi/2$. Let the particle move with velocity
$v=\tanh\x$, where $\x$ is the particle's Lorentz boost parameter. The
geometry is then sketched in Figure 1.

\midinsert\epsffile{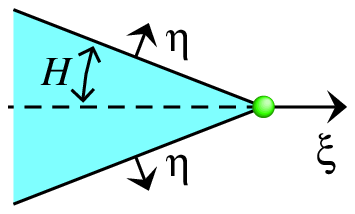}
\narrower\scrunch{Fig. 1. Wedge cut out by moving particle (dot). $\x$ 
is the boost parameter for the velocity of the particle, $\eta$ the one 
for the velocity of one edge of the wedge. The hamiltonian $H$ is half 
the wedge angle.}
\endinsert

By choosing the wedge such that the velocity vector lies in the
direction of its bisectrix one achieves that the jump across the wedge
is purely spacelike, and gluing the edges together provides with a
Cauchy surface. Lorentz contraction causes the wedge of a particle to
widen when it moves. The velocity of the edges, $\tanh\eta$, easily
follows from the geometry. The result is\ref{2,3,5,6}:

$$\eqalignno{\tan H\,=&\,\cosh\x\,\tan\m\,;&(2.1)\cr
\tanh\h\,=&\,\sin H\,\tanh\x\,;&(2.2)\cr
\cos\m\,=&\,\cos H\,\cosh\h\,;&(2.3)\cr
\sinh\h\,=&\,\sin\m\,\sinh\x\,.&(2.4)\cr}$$

\ni As explained in the introduction, since $\h$ is canonically conjugated 
to the length $L$ of the wedge, the hyperbolic sines and cosines 
suggest a discretization of the imaginary part of $L$, a somewhat 
mysterious feature.

When more particles are being considered the total Hamiltonian will
be the sum of the individual contributions, but there will also be 
``gravitational'' contributions from the vertex points between 
triples of polygons. This is because the Cauchy surface is curved at these 
points, so that the angles there do not add up to $2\pi$. These 
contributions are completely determined by the $\h$'s, as in Eq.~(2.3), 
but we will not discuss them further in this paper.

Now consider the coordinates $(x,y)$ of the particle. First let's have 
the particle move in the \hbox{$x$-direction}. Let $p_x$ be the 
canonical momentum associated to $x$. What is $p_x$? One must require 

$$\ddt x\ =\ \tanh\x\ =\ {\pa H\over\pa p_x}\,.\eqno(2.5)$$

\ni This is a partial derivative where $\m$ is kept fixed. The
Hamiltonian has been postulated to be half the wedge angle. We use Eq.
(2.3) to derive

$$\eqalign{\dd(\cos H\,\cosh\h)\,=\ 0\qquad\quad&\rightarrow\cr
\tan H\,\dd H\,=\,\tanh\h\,\dd\h\qquad&\rightarrow\cr \dd p_x\,=\,{\dd
H\over\tanh\x}\,=\,{\sin H\dd H\over\tanh\h}\,&=\cr =\,\cos
H\dd\h\,=\,{\cos\m\over\cosh\h}\dd\h\,&=\,(\cos\m)\dd\arctan(\sinh\h)
\,,\cr} \eqno(2.6)$$

\ni where also Eq.~(2.2) was used.

This we can write as

$$p_x\,=\,\th\cos\m\ ;\qquad \tan\th\,=\,\sinh\h\,,\eqno(2.7)$$

\ni which is synonymous to

$$\eqalignno{\sin\th\,=&\,\tanh\h\,;&(2.8)\cr
\cos\th\,\cosh&\h\,=\,1\,;&(2.9)\cr
\tan(\half\th)\,=&\,\tanh(\half\h)\,,&(2.10)\cr}$$

\ni provided we keep 

$$|\th|\,\le\,\pi/2\eqno(2.11)$$

\ni (a condition that we will ignore later). Thus we find that, apart
from a factor $\cos\m$, the canonical momentum $p_x$ is an angle $\th$.
\bigbreak

\ni{\bf 3. ENERGY-MOMENTUM SPACE, AND THE SPACE-TIME LATTICE}\medskip

Our first impression was that therefore the coordinate $x$ must be
quantized,

$$x\,\qu\,n_1/\cos\m\,,\eqno(3.1)$$

\ni and similarly the coordinate $y$. But then one hits two problems:
first, space would become a rectangular lattice, so that invariance
under continuous rotations would be lost, and furthermore, if the
particle does {\sl not\/} move in the $x$-direction, Eqs.
(2.6)---(2.11) would hold only for the absolute value $|p|$, not for
$p_x$ and $p_y$ separately.  We have to look at momentum space more
carefully. There are two angles:  the value $\th=p/\cos\m$, and the
angle $\vv$ describing the direction in which the particle moves, which
must be the direction of the vector $\bf p$. This, of course, is the
property of line segments lying on an $S_2$ sphere. Let us therefore
write

$$\eqalign{\tan(p_x/\cos\m)\,=&\,Q_1/Q_3\,;\cr
\tan(p_y/\cos\m)\,=&\,Q_2/Q_3\,.\cr}\eqno(3.2)$$

\ni Then $(Q_1,Q_2,Q_3)$ is a three-vector whose length is immaterial,  
whose angle with the 3-axis corresponds to the total momentum, and 
whose angle in the 1-2 plane is the angle of motion of the particle.

The space coordinates $x$ and $y$ should canonically
conjugated\fnd{Note however the remark at the end of Sect. 3.} to these
variables $p_x$ and $p_y$. Since $p_x$ and $p_y$ together form the
points of an $S_2$ sphere it is obvious what these conjugated variables
are: they are the quantum numbers $(\ell,m)$ of the spherical
harmonics. These indeed form a two-dimensional lattice, but one not
quite as simple as a rectangular one. Note that they allow for the full
rotation group $U(1)$ as well as discrete translations~--~the latter
being obtained by multiplying a spherical harmonic with another
spherical harmonic, using the Clebsch-Gordan coefficients for the
addition of angular momenta. How does this lattice have to be combined
with the timelike lattice, and how do we perform Lorentz
transformations?

The relation between the Hamiltonian and the momentum (that is, the 
Schr\"odinger equation), can be read off from Eqs. (2.9) and (2.3):

$$\cos H\,=\,\cos\m\,\cos\th\,.\eqno(3.3)$$

\ni Furthermore we have

$$\eqalignno{\tan\th\,=&\,\sin\m\,\sinh\x\,;&(3.4)\cr
\sin\th\,=&\,\sin H\,\tanh\x\,.&(3.5)\cr}$$

\ni These equations are to be compared with the usual relations for
non-gravitating relativistic particles (corresponding to the
$\m\rightarrow0$ limit):

$$\eqalignno{H^2\,=&\,\m^2+p^2\,;&(3.3a)\cr
p\,=&\,{\m v\over\sqrt{1-v^2}}\,;&(3.4a)\cr
p\,=&\,Hv\,.&(3.5a)\cr}$$

Let us now investigate how a Lorentz transformation in the $x$ direction 
affects $p_x$ and $H$. Let us increase the boost parameter $\x$ by a 
small Lorentz boost $\e$:

$$\eqalign{\x\,\rightarrow&\,\x+\e\,;\cr 
\sinh\x\,\rightarrow&\,\sinh\x+\e\cosh\x\,.\cr}\eqno(3.6)$$

\ni We find from Eq.~(2.7):

$$\eqalign{\tan\th\,\rightarrow&\,\tan\th+\e\sin\m\,\cosh\x \cr
&=\,\tan\th+\e(\cos\m)\tan H\,,\cr}\eqno(3.7)$$

\ni and, using Eq.~(2.1),

$$\eqalign{\tan H\,\rightarrow&\,\tan H+\e\tan\m\,\sinh\x\cr
&=\,\tan H+{\e\over\cos\m}\tan\th\,.\cr}\eqno(3.8)$$

\ni Clearly, the couple

$$\pmatrix{\tan\th\cr\cos\m\,\tan H\cr}\,,\eqno(3.9)$$

\ni which we can also write as 

$$\sin\m\pmatrix{\sinh\x\cr\cosh\x\cr}\,,\eqno(3.9a)$$

\ni transforms as a Lorentz vector. One may write 
 
$$\tan H\,=\,{Q_4\over Q_3}\,,\eqno(3.10)$$ 

\ni where $Q_4$ is again a {\sl real} quantity, so that 
we have invariance under the Lorentz transformation

$$\pmatrix{Q_4\cos\m\cr Q_1\cr Q_2\cr}\,\rightarrow\,L\,\pmatrix 
{Q_4\cos\m\cr Q_1\cr Q_2\cr}\,,\eqno(3.11)$$ 

\ni where $L$ is an ordinary $SO(2,1)$ matrix.  $Q_3$ is {\sl not\/}
involved in Lorentz transformations. There now are topologically
distinct ways to represent energy-momentum space. The length of the $Q$
vector can be chosen freely. Since we are dealing with angles, not
hyperbolic angles, it is not advised to use hyperbolic spaces but
rather compact spaces. If we take $Q1^2+Q_2^2+Q_3^2+Q_4^2=1$, this
space forms a Euclidean $S_3$ sphere. The $SO(4)$ spherical harmonics
then generate a lattice in space and time. Our first investigations
involved this lattice, and the field equations on it.  We now defer
further discussion of this lattice to the Appendix, because it was
discovered that the field equations one then obtains are difference
equations but they link {\sl five\/} consecutive time layers, which
prohibits the construction of any quantity that can serve as a
conserved, non-negative probability density. The square roots of these
equations, a discrete analogy of the Dirac equation, still links {\sl
three\/} consecutive time layers, which is still physically not
acceptable. We explain this situation further in the Appendix.

In our search for a physically most appealing theory we found that 
the $S_3$ topology for energy-momentum space must be abandoned, 
replacing it by the $S_2\times S_1$ torus. Thus we keep the $S_2$ 
sphere for momentum space and put the Hamiltonian on $S_1$.
In this energy-momentum space Lorentz transformations are still 
possible. We keep the 3-vector

$$\pmatrix{\cos\m\,\tan H\cr \tan\th\,\cos\vv\cr \tan\th\,\sin\vv\cr}
\eqno(3.12)$$

\ni (where $\vv$ is the angle in which the particle moves) as being the
one that transforms as usual under Lorentz transformations, but now
define points on the space-time lattice to correspond to harmonic
functions on this torus:

$$(\ell,m,t)\ \leftrightarrow\ Y_{\ell m}(\th,\vv)e^{-itH}\,,
\eqno(3.13)$$

\ni where $t$ is an integer denoting time. It is clear that here $\ell$
roughly corresponds to the distance from the origin\fndd{More
precisely:  $\ell(\ell+1)=(x^2+y^2)/\cos^2\!\m+m^2$.}, whereas $m$
exactly corresponds to angular momentum in the 2-plane. A discrete
Fourier transformation provides something like an angular position with
respect to the origin, so that the picture that emerges is as given in
Fig. 2a. Depicted there is the case when $\ell$ takes integral values.
Of course we also expect the possibility of half-integral values, in
which case the space lattice is as sketched in Fig. 2b. Since then
angular momentum has half-odd-integral values we expect this case to be
appropriate for fermions. We have not contemplated the possibility of
having `anyons' in this theory (see some remarks in the Discussion
Section).

\midinsert\epsffile{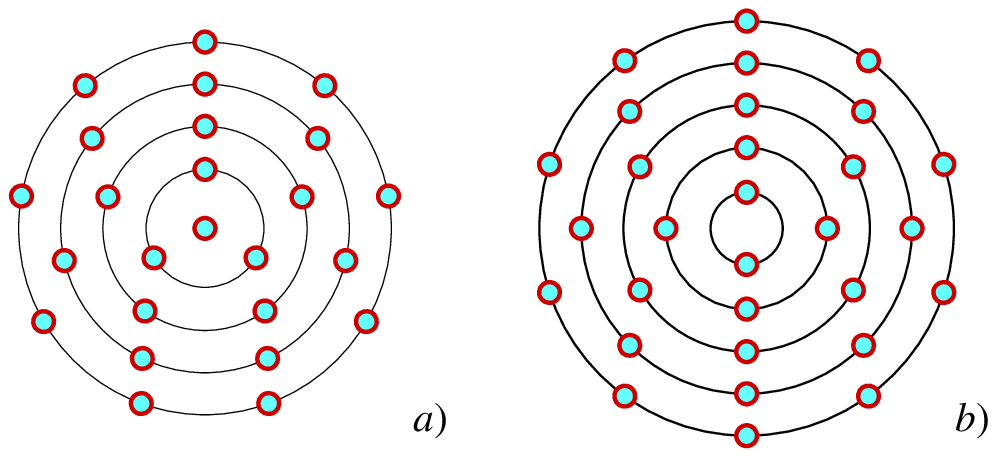}
\scrunch\cl{Fig.~2. Lattice in 2-space, a) the bosonic case, 
b) for fermions.}
\endinsert

It should be emphasized that a Lorentz transformation will not
correspond to a simple point transformation on our lattice. Lorentz
transformations do transform the harmonic functions into linear
superpositions of such functions. This implies that a Lorentz
transformation is not a local transformation in this theory. It is of
course a point transformation in energy-momentum space.

A second remark is that the coordinates $x$ and $y$ most closely
correspond to the angular momentum operators $L_2$ and $L_1$ on the
2-sphere:

$$x\,=\,{L_2\over\cos\m}\ ;\qquad y\,=\,{-L_1\over\cos\m}\ ;\qquad
L\,=\,L_3\,.\eqno(3.14)$$

\ni Consequently, we have the commutation rules

$$\eqalign{[x,y]\,=&\,{i\over\cos^{\,2}\!\m}L\,,\cr [L,x]\,=&\ iy\,,\cr
[L,y]\,=&\,-ix\,.\cr}\eqno(3.15)$$

\ni Here we reinserted the factors $\cos\m$ from Eq.~(2.7).

A third remark: the commutation rules between $\bf x$ and $\bf p$
appear to deviate from the standard expressions. This is because the
momenta have become angles. The commutation rules are now the ones
generated by the spherical harmonics $Y_{\ell m}$ in the usual way.
This is a delicate feature since we used Eq.~(2.5) as a starting
point.  It may be verified that Eq.~(2.5) holds in the limit of large
distances and large times, which is what one really must require for
the classical limit.

\bigbreak

\ni{\bf 4. A FIELD EQUATION}\medskip

Let us consider the 2-sphere

$$Q_1^2+Q_2^2+Q_3^2\,=\,1\,,\eqno(4.1)$$

\ni and define the momenta as in Eq.~(3.2). 
The mass shell condition, Eq.~(3.3), can then be rewritten as

$$e^{iH}+e^{-iH}\,=\,(2\cos\m)Q_3\,.\eqno(4.2)$$

\ni It is not difficult to see what this equation implies for functions 
defined on our space-time lattice. First consider {\sl any\/} function 
$\j(\ell,m,t)$. Since the functions $H$ and $Q_i$ are defined on its 
`Fourier transform' 

$$\hat\j(\th,\vv,H)\,=\,\sum_{\ell,m,t}Y^*_{\ell m}(\th,\vv)e^{iHt}
\j(\ell,m,t)\,,\eqno(4.3)$$ 

\ni the functions $e^{\pm iH}$ and $Q_i$ act as operators on $\j$.
The first cause a shift in time by one step down and up, respectively.
The action of the operators $Q_i$ is found as follows:

$$\eqalign{(Q_1+iQ_2)|\ell,\,m\ket\,&=\,C(\ell)\sqrt{(\ell+m+1)(\ell+m+2)}
\,|\ell+1,\,m+1\ket \cr
&-\,C(\ell-1)\sqrt{(\ell-m)(\ell-m-1)}\,|\ell-1,\,m+1\ket\,,}\eqno(4.4)$$
$$\eqalign{Q_3|\ell,m\ket\,&=\,C(\ell)\sqrt{(\ell+m+1)(\ell-m+1)}
\,|\ell+1,\,m\ket\cr 
&+\,C(\ell-1)\sqrt{(\ell+m)(\ell-m)}\,|\ell-1,\,m\ket\,,}\eqno(4.5)$$
$$\eqalign{(Q_1-iQ_2)|\ell,\,m\ket\,&=\,-C(\ell)\sqrt{(\ell-m+1)(\ell-m+2)}
\,|\ell+1,\,m-1\ket\cr
&+\,C(\ell-1)\sqrt{(\ell+m)(\ell+m-1)}\,|\ell-1,\,m-1\ket\,.}\eqno(4.6)$$

\ni where $C(\ell)$ is a normalization factor, depending only on $\ell$.
These equations are derived using standard Clebsch-Gordan techniques:
one considers the $Q$ functions in these equations as being
proportional to the states $|1,1\ket$, $|1,0\ket$ and $|1,-1\ket$,
respectively. The coefficients are then found by considering the
operators $L_\pm$ on these states. The normalization factor $C(\ell)$ is
found, after a series of algebraic manipulations, by inserting Eq.~(4.1):

$$C(\ell)\,=\,\Big((2\ell+1)(2\ell+3)\Big)^{-1/2}\,.\eqno(4.7)$$

Inserting Eq.~(4.5) into Eq.~(4.2) now yields a difference equation
which in the continuum limit should reproduce the Klein-Gordon
equation. Notice that this difference equation only involves nearest
neighbors in the $\{\ell,m,t\}$ lattice:

$$\eqalign{\j(\ell,m,t+1)\,+\,\j(\ell,m,t-1)\,&=\,(2\cos\m)\Bigg[
\sqrt{(\ell+m+1)(\ell-m+1)\over(2\ell+1)(2\ell+3)}\ \j(\ell+1,\,m,t)\cr
&+\,\sqrt{(\ell+m)(\ell-m)\over(2\ell-1)(2\ell+1)}\ \j(\ell-1,\,m,t) 
\Bigg]\,.\cr}\eqno(4.8)$$

\ni For large $\ell$, small $m$ and small $\m$ this approaches to:

$$\eqalign{\j(\ell,m,t+1)-2\j(\ell,m,t)+\j(\ell,m,t-1)\ =\ & 
 {2-\m^2\over2}\Big[1-{m^2\over2\ell^2}+{1\over8\ell^2}\Big]\,\times\cr
\Big[\j(\ell+1,m,t)-2\j(\ell,m,t)+\j(\ell-1,m,t)\Big]&-\Big[\m^2+
{m^2-\quart\over\ell^2}\Big]\j(\ell,m,t)\,.\cr}\eqno(4.9)$$

\ni Identifying here $m$ with angular momentum and $\ell$ with 
$r=\sqrt{x^2+y^2}$, we recognise the Klein-Gordon equation in polar 
coordinates.

The fact that Eq.~(4.8) emerges as a difference equation involving only
nearest neighbors is the primary physical reason for introducing the
$\{\ell,m,t\}$ lattice as proposed in this paper.\bigbreak

\ni{\bf 5. THE LATTICE DIRAC EQUATION}\medskip

Since Eq.~(4.8) connects three time slices it is not suitable for 
serving as a quantum wave function. We believe that a promising route 
towards a useful quantized model for gravitating particles in 2+1 
dimensions is to follow Dirac's philosophy: find an acceptable wave 
function for fermions, which allows for an interpretation of $|\j|^2$ 
as a probability distribution, after which we handle the negative 
energy solutions by filling all these levels. This paper will not go 
that far (there are various important problems that would yet have to 
be addressed), but the first step, a Lorentz covariant equation 
with a non negative preserved probability distribution, can now be
made.

Instead of Eq.~(4.2) we need an equation containing $e^{-iH}$ only,
so that an evolution operator would be obtained connecting exactly 
two time slices. To do this, we need not only an expression for 
$\cos H$, as in Eq.~(3.3), but also one for $\sin H$. We have, on
the sphere (4.1):

$$\eqalign{\sin^2\!H\,=&\,\sin^2\!\m+\cos^2\!\m\,\sin^2\!\th\,=\cr 
=&\,\sin^2\!\m+\cos^2\!\m\,(Q_1^2+Q_2^2)\,.\cr}\eqno(5.1)$$

\ni Therefore

$$\eqalign{e^{-iH}\,&=\,\cos\m\,Q_3\,-i\sqrt{\sin^2\!\m+
\cos^2\!\m\,(Q_1^2+Q_2^2)}\,=\cr &=\,\cos\m\,Q_3\,-i\sqrt
{Q_1^2+Q_2^2+\sin^2\!\m\,Q_3^2}\,.\cr}\eqno(5.2)$$

\ni To have the equation homogeneous in the $Q$'s will be of importance
later. Clearly, the square root in this equation would turn it into a
non-local one on the lattice. Following Dirac we replace this equation
by a linear one:

$$e^{-iH}\j\,=\,\big[\cos\m\,Q_3-i\a_1Q_1-i\a_2Q_2-i\b\sin\m\,Q_3\big]
\j\,,\eqno(5.3)$$

\ni where $\a_i$ and $\b$ obey the usual Dirac anticommutation rules,
so that the Eigenvalues of the operator in (5.3) correspond to the
field equation (4.8).

To check Lorentz invariance we consider the Lorentz vector (3.11), 
(3.12).\hfil\break $Q_4$, as defined in (3.10), obeys

$$Q_3^2+Q_4^2\,=\,1/\cos^2\!\m\,,\eqno(5.4)$$

\ni so that

$$e^{-iH}\,=\,\cos\m\,(Q_3-iQ_4)\,,\eqno(5.5)$$ 

\ni and Eq.~(5.3) can be rewritten as 

$$\cos\m\,Q_4\,=\,\a_1Q_1+\a_2Q_2+\b\sin\m\,Q_3\,,\eqno(5.6)$$

\ni or, renaming $\b$ as $\g_3$ and $\b\a_i$ by $i\g_i$, we get 

$$\cos\m\,\g_3Q_4\,=\,i\g_1Q_1+i\g_2Q_2+\sin\m\,Q_3\ ;\qquad 
\{\g_i,\g_j\}\,=\,2\d_{ij}\,.\eqno(5.7)$$

\ni This equation of course has the required Lorentz covariance if
(3.11) transforms as a Lorentz vector. Note that without the second 
step in Eq.~(5.2) we would have obtained a different equation that 
does not appear to have the required invariance.

Let us rewrite Eq.~(5.3) as

$$e^{-iH}\j\,=\,\big[e^{-i\b\m}Q_3-i\a_iQ_i\big]\j\,.\eqno(5.8)$$

\ni On the lattice it becomes

$$\eqalign{\j(\ell,m,t+1)\,&=\,e^{-i\g_3\m}\big[F^1_{\ell,m}
\j(\ell+1,m,t) +F^2_{\ell,m}\j(\ell-1,m,t)\big]\,+\cr &+\,\g_3
{\g_1-i\g_2\over2}\big[ F^3_{\ell,m}\j(\ell+1,m+1,t)-F^4_
{\ell,m}\j(\ell-1,m+1,t)\big]\,+\cr &+\,\g_3{\g_1+i\g_2\over2}
\big[-\!F^5_{\ell,m}\j(\ell+1,m-1,t)+F^6_{\ell,m} \j(\ell-1,m-1,t)
\big]\,.\cr}\eqno(5.9)$$

\ni where the coefficients can be read off from Eqs (4.4)--(4.7):

$$\eqalign{F^1_{\ell,m}\,=\,\sqrt{(\ell+m+1)(\ell-m+1)\over(2\ell+1)
(2\ell+3)}\,,\qquad& F^2_{\ell,m}\,=\,\sqrt{(\ell+m)(\ell-m)
\over(2\ell-1)(2\ell+1)}\,,\cr F^3_{\ell,m}\,=\,\sqrt{(\ell+m+1)
(\ell+m+2)\over(2\ell+1)(2\ell+3)}\,, \qquad& F^4_{\ell,m}\,
=\,\sqrt{(\ell-m)(\ell-m-1)\over(2\ell-1)(2\ell+1)} \,,\cr
F^5_{\ell,m}\,=\,\sqrt{(\ell-m+1)(\ell-m+2)\over(2\ell+1)
(2\ell+3)}\,,\qquad& F^6_{\ell,m}\,=\,\sqrt{(\ell+m)(\ell+m-1)\over
(2\ell-1)(2\ell+1)}\,.\cr}\eqno(5.10)$$

By construction, our equation (5.8) produces Eigenvalues with absolute 
values equal to one, and hence we immediately see that the norm of 
$\j$ is preserved:

$$\|\j(t+1)\|\,=\,\|\j(t)\|\,.\eqno(5.11)$$

\ni This equation can be used as an evolution equation for a
quantum state $\j$. Needless to say, our Hamiltonian $H$ has as many
positive Eigenvalues as negative ones, and so an eventual physically
viable theory should have the negative states filled up like a Dirac
sea, or else it will be impossible to perform thermodynamical
calculations. As stated before, this would go beyond the scope of the
present paper.
\bigbreak

\ni{\bf 6. DISCUSSION}\medskip

The introduction of lattice versions of gravity has been proposed 
before\ref8. This particuar lattice may also have been considered before.
Here we emphasize that the lattice structure has been {\sl derived\/} 
from the mathematical expressions for the relation between momenta 
and the Hamiltonian, rather than postulated. Only with the lattice 
spacing precisely defined by the Planck length do the field equations 
take the form of difference equations involving only nearest 
neighbors on the lattice.

It should be kept in mind that the lattice structure is more
complicated than the familiar rectangular or triangular lattices. For
a given value of $\ell$, the angular momentum is limited by the
inequality

$$|m|\,\le\,\ell\,.\eqno(6.1)$$
 
\ni The distance to the origin, $r$, is quantized as 

$$r\,=\,{1\over\cos\m}\sqrt{\ell(\ell+1)-m^2}\,,\eqno(6.1)$$

Concerning the quantum of angular momentum, the reader might have
wondered why it should be taken as integers (or half-odd-integers) at
all. First of all, one could consider ``anyon'' statistics. Now this
appears to be difficult in our approach. If energy-momentum space is
indeed represented by an $S_2$ sphere we will be forced to restrict
ourselves to the $SO(3)$ quantum numbers $\ell$ and $m$, but it is
conceivable that more complex theories can be constructed; we leave
this to other investigators.

Secondly however, one might argue that we also have to deal with the
non-trivial holonomies generated by the cusps around the particles. It
may seem that angular momentum ought to be quantized in units
$\pi/(\pi-H)$ rather than 1. We deliberately postponed this
complication.  The reason is that this topological aspect of space
emerges if {\sl one particle circumnavigates another particle}, in
other words, it becomes relevant only if we consider two or more
particles. In the present work only single particle states were
considered. The center of our coordinate frame is not a particle. It
cannot be taken as an infinitely heavy particle because such a thing
would distort the surrounding space too much. As yet we treat the
origin as a more abstract reference point in our coordinates. This
situation is clearly not completely satisfactory. In order to construct
a more complete theory one will have to addres the many particle
situation.\fnd{An approach towards handling this problem using
conformal gauges for space and time is advocated by Menotti,
Seminara\ref9 and Welling\ref{10}.}  One then can restrict oneself to
describing distances between particles only, while avoiding abstract
reference points altogether. This however will be a major exercise,
complicated by the fact that in the polygon representation (which we
insist on using since it involves proper Cauchy surfaces) shows extra
contributions to the Hamiltonian from the vertex points between the
polygons. We suspect that such a treatment will also further clarify
the role of the factors $\cos\m$ in our expressions (see Eq.~6.1),
which presently seem to be rather ugly.

The equations on our lattice are local in the sense that only nearest
neighbors interact, and the Cauchy surfaces are well-defined. In
contrast, Poincar\'e transformations are in general non-local. Related
to this is the fact that we have some freedom in choosing the details
of our lattice. Instead of $S_2\times S_1$ we could have picked
$S_3$.  What happens then is explained in the Appendix. The equations
also involve nearest neighbors, except that in the time direction
next-to-nearest neighbors then also appear, and this makes this
lattice less suitable. The arguments of Sect. 5, in particular Eq.
(5.11), do not work in that case.

The Poincar\'e group in our model is different from the usual one. We
still have the complete Lorentz group, but the translations are limited
to the procedures of adding angular momenta using the familiar
Clebsch-Gordan coefficients. 

We seem to have hit upon a new terrain of lattice theories where much 
remains to be explored.

\bigskip\ni{\bf ACKNOWLEDGEMENT}\medskip

The author thanks T. Jacobson, M. Ortiz and M. Welling for valuable
discussions on this topic.\bigbreak

\ni{\bf NOTE ADDED}\medskip

Just before this manuscript was mailed a paper by Waelbroeck and Zapata
appeared in the electronic archives\ref{11} in which also a space-time
lattice is discussed in connection with 2+1 dimensional gravity. His
lattice however is the lattice formed by the imaginary parts of the
link variables $L_i$ as discussed in the Introduction. It is quite
distinct from what we propose in this paper.

\bigbreak
{\ni \bf APPENDIX: THE LATTICE GENERATED BY $S_3$.}\medskip

The algebra of the spherical harmonics on $S_3$ is generated by first
considering rotation operators $L_a$ and $M_a$, $\,a=1,2,3$, in a
four-dimensional space with coordinates $(q_1,\dots,q_4)$:

$$L_a\,=\,-i\e_{abc}q_b\pa_c\ ;\qquad M_a\,=\,i(q_4\pa_a-q_a\pa_4)\,.
\eqno(A.1)$$

\ni They obey the commutation rules

$$\eqalignno{[L_a,L_b]\,=&\,i\e_{abc}L_c\,;&(A.2a)\cr
[L_a,M_b]\,=&\,i\e_{abc}M_c\,;&(A.2b)\cr
[M_a,M_b]\,=&\,i\e_{abc}L_c\,.&(A.2c)\cr}$$

\ni Construct the self-dual and the anti-self-dual combinations,

$$L^L_a\,=\,\half(L_a+M_a)\ ;\qquad L^R_a\,=\,\half(L_a-M_a)\,. 
\eqno(A.3)$$ 

\ni We have 

$$[L^L_a,L^L_b]\,=\,i\e_{abc}L^L_c\ ;\qquad [L^R_a,L^R_b]\,=\,i\e_{abc}
L^R_c\ ;\qquad [L^L_a,L^R_b]\,=\,0\,.\eqno(A.4)$$

\ni Since ${\bf L\cdot M}=0$ we find

$$\eqalign{{{\bf L}^L}^2\,=\,{{\bf L}^R}^2\,=&\,\quart({\bf L}^2+
{\bf M}^2)\,=\cr =&\,-q_\m^2\pa_\n^2+(q_\m\pa_\m)^2+2q_\m\pa_\m\,
=\,\ell(\ell+1)\,.\cr}\eqno(A.5)$$

\ni The simplest non-trivial function is $f_\m(q)\,=\,q_\m$, which has 
$$\pa_\m^2f_\n\,=\,0\ ;\qquad(q_\m\pa_\m )f_\n\,=\,f_\n\,,\eqno(A.6)$$ 

\ni so that we read off from Eq.~(A.5) that it has $\ell=\half$. Its
four components are represented by $m_L=\pm\half,\  m_R=\pm\half$.
These functions $f_\m$ are then to be considered on the sphere
$|q|=1$.  By multiplying these together we can construct functions with
all other (integral or half integral) $\ell$ values, whereas on the
other hand it is easy to convince oneself that every set of numbers
$(\ell,m_L,m_R)$ with $|m_L|\le\ell$ and $|m_R|\le\ell$ corresponds to
one and only one spherical harmonic function (up to a phase factor).
These numbers $\ell$, $m_L$ and $m_R$ therefore form a complete set of
coordinates. 

We have a freedom in our choice for orienting the coordinates $Q_\m$ in
the $q_\m$ space. How one does this turns out to make little
difference; in all cases the $\ell,\,m_L,\,m_R$ do not directly
correspond to all space-time coordinates. A convenient choice is:

$$Q_1=q_1\,,\quad Q_2=q_2\,,\quad Q_3=q_4\,,\quad Q_4=q_3\,.
\eqno(A.7)$$
 
\ni Then the variable conjugated to the Hamiltonian (3.10) is

$$t\,=\,-M_3\,=\,m_R-m_L\,.\eqno(A.8)$$

\ni and from Eqs. (3.2) and (A.1) we have

$$x\cos\m\,=\,L_2\ ;\qquad y\cos\m\,=\,-L_1\,,\eqno(A.9)$$

Consider an {\sl arbitrary\/} function 
of space-time, {\sl i.e.}, one that does not obey any equation of motion 
such as a mass shell condition. For such a function the three coordinates 
$x$, $y$, and $t$ can all three be regarded as operators. They are 
quantized according to

$$x_i\,=\,n_i/\cos\m\ ;\qquad t\,=\,n_3/\cos\m\,,\eqno(A.10)$$ 

\ni however, these three numbers $n_i$ are non-commuting. We have

$$\eqalign{[x,y]\,=\,i(\cos^2\!\m)L_3\,,\cr
[y,t]\,=\,-i(\cos\m)M_2\,,\cr
[x,t]\,=\,-i(\cos\m)M_1\,.\cr}\eqno(A.11)$$ 

\ni The operators $L_{1,2,3}$ are the usual generators of {\sl
Euclidean} rotations among the $M_i$. Thus only the operators $t$,
$L_3$, and $ \ell$ are diagonalized simultaneously. $\ell$ roughly
corresponds to `total Euclidean distance from the origin'. If we would
limit ourselves to integral $\ell$ we find that at even time $m_R-m_L$
is even, therefore $m_R+m_L$ is even.  This would imply that at even
time {\sl parity} of all functions $f(x,y)$ is even: $f(x,y)=f(-x,-y)$,
and at odd times parity is odd. If one wants to avoid that,
half-odd-integer values of $\ell$ should be admitted together with the
integral ones.

The $L_i$ represent Euclidean rotations in space-time.  Lorentz
invariance is not a property of the coordinates but of the mass shell
condition (3.3). Lorentz transformations are defined in terms of
Eq.~(3.11) in energy-momentum space. They map the sphere onto itself,
after which decomposition in spherical harmonics will give the new
coordinates.  Since the $L$'s and $M$'s of Eqs (A.2) are the generators
of Euclidean rotations they do not transform into combinations of each
other under Lorentz transformations. Therefore the Lorentz group will
act in a fairly complicated way on our coordinates; this statement
holds both for the $S_3$ case and for the $S_2\times S_1$ case treated
in the text.

\midinsert\epsffile{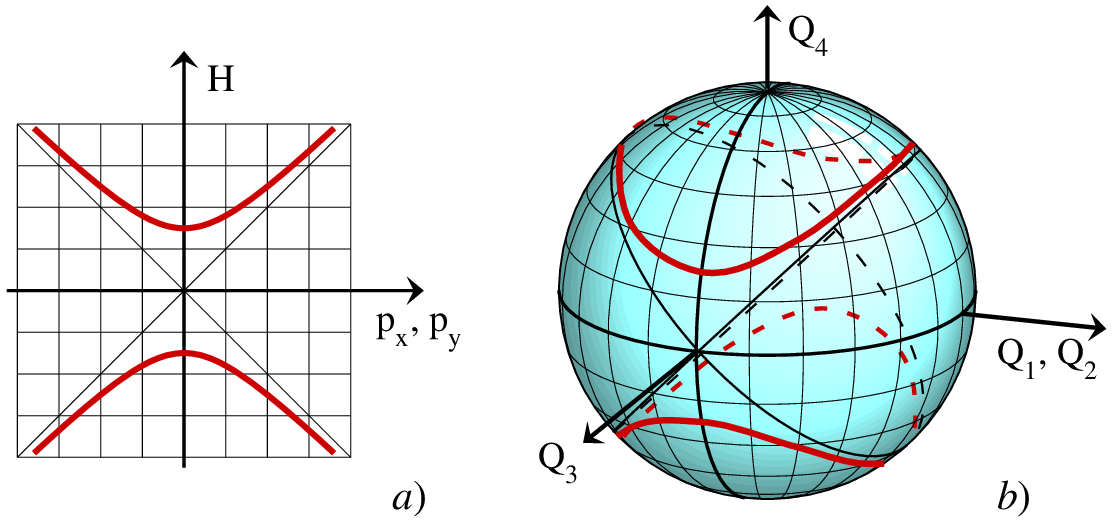} \scrunch\narrower{Fig.~3. The mass
shell: a) in ordinary 2+1 dimensional particle theory, b) in 2+1
dimensional gravity for the $S_3$ case. Energy and momentum live on an
$S_3$ sphere. In both cases the two space axes are drawn as one.}
\endinsert

Now consider the mass shell condition, Eq.~(3.3), which in the
$S_2\times S_1$ case had to be rewritten as Eq.~(4.2). In order to be
able to process it in $S_3$ we have to rewrite it as

$$Q_4^2\cos^2\!\m\,=\,Q_1^2+Q_2^2+Q_3^2\sin^2\!\m\eqno(A.12)$$

\ni (compare Eq.~(5.2)). In ordinary flat space
it is replaced by Eq.~(3.3a), which corresponds to the Klein-Gordon
equation in coordinate space. In Fig.~3 we show how the mass shell looks 
on the $S_3$ sphere. What is the equation generated by (A.12)
in our space-time lattice? The functions $Q_\m Q_\n$ form a
representation of $SO(4)$ obtained from the symmetrized product of two
$\ell=\half$ representations $Q_\m$:

$$4\times5\ /\ 2\ =\ 10\ =\ 9+1\,,\eqno(A.13)$$ 

\ni where the 1 is of course the scalar $Q_\m^2$, which is fixed to be
1 on the sphere.  The 9 is formed from the states $\ell=1$,
$|m_L|\le1$, $|m_R|\le1$.  Multiplying a function at the point
$\ell,m_L,m_R$ with this representation leads one to a neighboring
point $\ell+\D\ell,\, m_L+\D m_L,\, m_R+\D m_R$, with $|\D\ell|\le 1$,
$\D m_L|\le 1$ and $|\D m_R|\le 1$. We see that, as in Sect. 4, the
Klein-Gordon equation is replaced by a difference equation on the
lattice. We also see that only points differing by integral amounts of
$\ell$, $m_L$ and $m_R$ are connected, so that the space-time lattice
breaks up into two mutually non-interacting systems. This corresponds
to the conservation of parity.

Let us consider the action of the $Q_\m$ operators (or equivalently the 
$q_\m$) in somewhat more detail. From Eqs. (A.1) and (A.3) we have 

$${L_3}^{L\atop R}\,=\,\half i(q_2\pa_1-q_1\pa_2\pm q_4\pa_3\mp q_3\pa_4) 
\,.\eqno(A.14)$$

\ni In order to expand a ``state'' $\sum_\m \a_\m q_\m$ in terms of
eigenstates $|\half,\,m_L,\,m_R\ket$, we solve the equations

$$\eqalign{{L_3}^{L\atop R}(\a\cdot q)\,&=\,\half i(q_2\a_1-q_1\a_2\pm
q_4\a_3\mp q_3\a_4)\cr&=\,m_{L\atop R}(\a_1 q_1+\a_2 q_2+\a_3 q_3+ \a_4
q_4)\,, \cr}\eqno(A.15)$$

\ni Indicating the states $|\half,m_L,m_R\ket$ (where of course all $m$
values are $\pm\half$) by the short-hand notation $|\s_L,\s_R\ket$,
where the $\s$'s denote the signs of the $m$'s, we find:
\def\srt{\sqrt2\,}

$$\matrix{&&\a_1&\a_2&\a_3&\a_4\cr \srt|++\ket&\ra&1&i&0&0\cr 
\srt|+-\ket&\ra&0&0&-1&-i\cr \srt|-+\ket&\ra&0&0&-1&i\cr
\srt|--\ket&\ra&-1&i&0&0\cr} \qquad\qquad\matrix{&\vphantom{a_1}&\cr
\srt q_1&=&|++\ket\,-\,|--\ket\cr
\srt q_2&=&-i|++\ket\,-\,i|--\ket\cr \srt q_3&=&-|+-\ket\,-\,
|-+\ket\cr \srt q_4&=&i|+-\ket\,-\,i|-+\ket\cr}\eqno(A.16)$$

\ni The relative signs here have been fixed by considering the (left 
and right) angular momentum raising and lowering operators.

Now that we know their left and right quantum numbers we are in a
position to calculate the action of the operators $q_\m$ on the states
$|\ell,m_L,m_R\ket$. The $m$ quantum numbers just add up, whereas the
total angular momentum numbers $\ell$ change by one half unit. Thus we
can state that

$$\eqalign{|\s_L,\s_R\ket|\ell,m_L,m_R\ket\,=&\,\a^{\s_L,\s_R}_{m_L,m_R}
|\ell+\half, \,m_L+\half\s_L,\,m_R+\half\s_R\ket\,+\cr
+&\,\b^{\s_L,\s_R}_{m_L,m_R}|\ell-\half,
\,m_L+\half\s_L,\,m_R+\half\s_R\ket\,,\cr}\eqno(A.17)$$

\ni where $\a$ and $\b$ are coefficients that can now be calculated. The
technique to be used is a standard manipulation with Clebsch-Gordan
coefficients, similar to what we did in Sect. 4, but we will not repeat
it here.
 
We see confirmed that the equation (A.12), in which each term
corresponds to acting twice with operations as described in Eq.~(A.17),
amounts to a difference equation on the $(\ell,m_L,m_R)$ lattice.  In
general this equation takes the form:

$$\sum_{\s_\ell,\s_L,\s_R}F(\ell,m_L,m_R,\s_\ell,\s_L,\s_R)\ 
\j(\ell+\s_\ell,\,m_L+\s_L,\,m_R+\s_R)\,=\,0\,,\eqno(A.18)$$
 
\ni where all variables $\s_{\ell,L,R}\,$ take the values $\pm1$ only.
Now, in view of Eq.~(A.8), we see that to describe the complete space
of solutions to this equation one needs the Cauchy data on several
consecutive ``Cauchy planes'': $t=t_1+1,\ t_1,\dots$, $t_1-2$,
after which the values at $t=t_1+2$ follow.

The Dirac equation again takes the form 

$$i\g_1 Q_1+i\g_2Q_2+\g_3Q_4\cos\m+Q_3\sin\m\,=\,0\,,\eqno(A.19)$$

\ni where $\g_{1,2,3}$ can be taken to be the three Pauli matrices. 
Multiplying the equation with 

$$i\g_1 Q_1+i\g_2Q_2+\g_3Q_4\cos\m-Q_3\sin\m\,,\eqno(A.20)$$

reproduces (A.12). Note that Lorentz covariance follows from the usual 
arguments, using the transformation law (3.11).

Consider a solution of the difference equation (A.19). In the notation
of Eq.~(A.16) our equation (A.19) takes the form

$$(\g_2+i\g_1)|++\ket+|\g_2-i\g_1)|--\ket+i\g_3e^{i\g_3\m}|+-\ket-
i\g_3e^{-i\g_3\m} |-+\ket\,=\,0\,.\eqno(A.21)$$

\ni Since $t=m_R-m_L$ (Eq.~A.8), the last term advances time by one
unit and the third term retards by one unit. The first two terms stay
on the same time sheet.  One can never have a positive and conserved
probability density for such a wave equation.

In conclusion, besides giving mathematically more cumbersome difference
equations, the $S_3$ lattice is also physically less appealing than
$S_2\times S_1$. Ultimately, however, both physical systems are
mathematically equivalent, since in momentum space the field equations
are identical. \bigbreak

\ni{\bf REFERENCES}\medskip
\def\br{\hfil\break}

\item{1.}  J.R. Gott and M. Alpert, {\it Gen. Rel. Grav.} {\bf 16}
(1984) 243; \br S. Giddings, J. Abbot and K. Kuchar, {\it Gen. Rel.
Grav.} {\bf 16} (1984) 751;\br J.E. Nelson and T. Regge, {\it
Quantisation of 2+1 gravity for genus 2}, Torino prepr. DFTT 54/93,
gr-qc/9311029

\item{2.} A. Staruszkiewicz, {\it Acta Phys. Polon}. {\bf 24} (1963)
734; \br J.R. Gott, and M. Alpert, {\it Gen. Rel. Grav.} {\bf 16}
(1984) 243; \br S.  Giddings,  J. Abbot and  K. Kuchar, {\it Gen. Rel.
Grav.} {\bf 16} (1984) 751; \br S. Deser, R. Jackiw and G. 't Hooft,
{\it Ann. Phys.} {\bf 152} (1984) 220; iid, Phys. Rev. Lett. {\bf 68}
(1992) 267.

\item{3.} S.M. Carroll, E. Farhi and A.H. Guth, {\it Phys. Rev. Lett.}
{\bf 68} (1992) 263, \br C. Cutler, {\it Phys. Rev.} {\bf  D 45} (1992)
487, see also A.  Ori, {\it Phys.  Rev.}  {\bf D44} (1991) R2214; \br
G. 't Hooft, {\it Class.  Quantum Grav.} {\bf 10} (1993) 1023, {\it
ibid.} {\bf 10} (1993) S79; {\it Nucl. Phys.} {\bf B30} (Proc. Suppl.)
(1993) 200.

\item{4.}  E. Witten, {\it Nucl. Phys.} {\bf B311} (1988) 46; \br S.
Carlip, {\it Nucl. Phys.} {\bf B324} (1989) 106; and in: ``{\it
Physics,  Geometry and Topology\/}", NATO ASI series B, Physics, Vol.
{\bf 238}, H.C.  Lee ed., Plenum 1990, p. 541; \br S. Carlip, {\it Six
ways to quantize (2+1)-dimensional gravity}, Davis Preprint UCD-93-15,
gr-qc/9305020. \br G. Grignani, {\it 2+1-dimensional gravity as a gauge
theory of the Poincar\'e group}, Scuola Normale Superiore, Perugia,
Thesis 1992-1993; \br G. 't Hooft, {\it Commun. Math. Phys.} {\bf 117}
(1988) 685;  S. Deser and R.  Jackiw, {\it Comm. Math. Phys.} {\bf 118}
(1988) 495.

\item{5.} G. 't Hooft, {\it Class. Quantum Grav.} {\bf 10} (1993) 1653
(gr-qc/9305008).\br See also: A.P. Balachandran and L. Chandar, {\it
Nucl. Phys.} {\bf B 428} (1994) 435.

\item{6.} G. 't Hooft, {\it Class. Quantum Grav.} {\bf 9} (1992) 1335.

\item{7.} H. Waelbroeck, {\it Class. Quantum Grav.} {\bf 7} (1990) 751;
{\it Phys. Rev. Lett.} {\bf 64} (1990) 2222; {\it Nucl. Phys.} {\bf B
364} (1991) 475; H.~Waelbroeck and F.~Zertuche, {\it Phys. Rev.} {\bf
D50} (1994) 4966

\item{8.} See for instance: T. Regge, {\it Nuovo Cimento} {\bf 19}
(1961) 558; \br J. Ambj\o rn and J. Jurkiewicz, {\it Phys. Lett.} {\bf
B 278} (92) 42;\br J. Ambj\o rn, J. Jurkiewicz, and C.F. Kristjansen,
{\it Nucl. Phys. \bf B 393} (93) 601;\br M.E. Agishtein and A.A.
Migdal, {\it Mod.  Phys. Lett. \bf A7} (92) 1039; {\it Nucl. Phys. \bf
B385} (92) 395. \br G.~'t Hooft, in ``{\it Recent Developments in
Gravitation\/}", Carg\'ese 1978, ed. M. L\'evy and S. Deser, Plenum,
New York/London, p. 323; \br B.V.  de Bakker, and J. Smit, {\it Phys.
Lett. \bf B334} (1994) 304; {\it Nucl. Phys. \bf B439 }(1995) 239.

\item{9.}   P. Menotti and D. Seminara, {\it Fermi-Walker gauge in 2+1
dimensional gravity}, Pisa/MIT preprint MIT-CTP\#2400, IFUP-TH-1/95,
January 1995.

\item{10.} M. Welling, {\it Gravity in 2+1 Dimensions as a
Riemann-Hilbert Problem}, Utrecht prepr. THU~95/24 (hep-th/9510060);
id, {\it Some Approach to 2+1 Dimensional Gravity Coupled to Point
Particles}, Utrecht prepr. THU~95/33 (hep-th/9511211).

\item{11.} H.~Waelbroeck and J.A.~Zapata, {\it 2+1 Covariant Lattice
Theory and 't~Hooft's Formulation}, Pennsylvania State Univ. prepr.
CGPG-95/8 (gr-qc/9601011).

\bye